\documentclass[aps,pre,reprint,showkeys]{revtex4-1}
\usepackage{graphicx}
\usepackage{amssymb}
\usepackage{epstopdf}
\usepackage{amsmath}

\newcommand{\im}{\text{Im\,}}
\newcommand{\re}{\text{Re\,}}
\newcommand{\bigt}{\mathbf{T}}
\newcommand{\osc}{\mathcal{ Q}}
\newcommand{\lencycle}{\ell_{cyc}}

\newcommand{\longrightharpoonup}{\relbar\joinrel\rightharpoonup}
\newcommand{\longleftharpoondown}{\leftharpoondown\joinrel\relbar}

\newcommand{\longrightleftharpoons}{
  \mathop{
    \vcenter{
      \hbox{
      \ooalign{
        \raise1pt\hbox{$\longrightharpoonup\joinrel$}\crcr
	  \lower1pt\hbox{$\longleftharpoondown\joinrel$}
	  }
      }
    }
  }
}

\begin{document}

\title{The Quality of Oscillations in Overdamped Networks}
\author{Nathan Hodas }
\email[]{nhodas@isi.edu}
\affiliation{Beckman Institute, Biological Imaging Center}
\affiliation{ Dept. of Physics, Caltech Pasadena, CA 91125}

\keywords{oscillations, nonequilibrium statistical dynamics, limit-cycles, molecular motors, network dynamics}

\date{\today}                                           

\begin{abstract}
The second law of thermodynamics implies that no macroscopic system may oscillate indefinitely without consuming energy.  The question of the number of possible oscillations and the coherent quality of these oscillations remain unanswered.  This paper proves the upper-bounds on the number and quality of such oscillations when the system in question is homogeneously driven and has a discrete network of states.  In a closed system, the maximum number of oscillations is bounded by the number of states in the network.  In open systems, the size of the network bounds the quality factor of oscillation.  This work also explores how the quality factor of macrostate oscillations, such as would be observed in chemical reactions, are bounded by the smallest equivalent loop of the network, not the size of the entire system. The consequences of this limit are explored in the context of chemical clocks and limit cycles. 
\end{abstract}

\maketitle

\section{Introduction}
Oscillations are ubiquitous.  We celebrate them and attempt to harness them.  Naturally, this interest drives us to study them. There has been no shortage of analysis of the simple harmonic oscillator in all of its variations, but much of the periodicity around us is not equivalent to a mass on a spring. For example,  the beating heart is driven by molecular motors which exist in the low-Reynolds number regime, where viscous damping overwhelms inertia. For these molecular constituents, buffeting by solvent molecules prevents coherent oscillations from persisting on a timescale longer than the mean time between collisions, which is on order picoseconds~\cite{coffey2004langevin}.  Despite this, we observe the coordination of overdamped components to produce periodic behavior~\cite{Kay:2007p1230}.  Studying this coordination on a problem-by-problem basis has uncovered some conceptual principles for creating oscillatory behavior in the overdamped regime, but few truly fundamental laws exist~\cite{FIELD:1989p1191,Placais:2009p1199,Gao:2003p578,Helbing:2004p530,Green:2008p1231}.   This work bounds the  oscillatory performance of all discrete-state overdamped systems, providing a new look at the necessary conditions for creating coherent oscillations in dynamic networks~\footnote{We all possess an intuitive comfort with oscillations, but we have to formalize this notion for our analysis. To separate coherent oscillations from random fluctuations, we demand that oscillations be predictable and have a characteristic timescale. Predictability implies that the autocorrelation of a signal will have distinct peaks or troughs corresponding to the period of the oscillations.}.

 When the energy landscape of an overdamped system can be divided into distinct basins of attraction with barriers higher than \(k_BT\), the system will tend to reach a local equilibrium within a basin before fluctuations stochastically drive it over a barrier into a neighboring basin.  When this condition holds, it is common and appropriate to model each basin as a distinct state, with a static rate of transitioning from one state to another~\cite{Svoboda:1994p1220,Lipowsky:2008p1238,Linden:2008p1242}, meaning the entire system may be represented as network with nodes representing each state and a edge connecting nodes with a non-zero transition rate. 
 These systems are finite state first-order Markov processes~\footnote{This means we exclude systems with high degrees of quantum coherence and also those that are underdamped and therefore inertial.  This means the systems of interest  arrive at local equilibrium  before changing states.} and can be modeled by the finite-state master equation:
\begin{equation}\label{master_equation}
\frac{d p_i (t)}{dt} = \sum_j^N T_{ij} p_j(t)  -\sum_{j \neq i}^N T_{ij} p_i(t),
\end{equation}
where \(T_{ij}\) is the transition rate from state \(j\) to state \(i\).  For introductions to the master equation and its numerous  applications, see~\cite{van2007stochastic,Schmiedl:2007hw,gardiner2004handbook,Helbing:2010vq}. We will assume that all rates are time-independent, meaning no external factors change the rates (but this does not necessarily mean that the system is closed).  We also make the assumption that \(\bigt\) is an irreducible matrix, enforcing the trivial condition that we are not modeling multiple mutually isolated systems.  Finally, we assume that the systems conserve probability, which can always be enforced by adding states to the system to represent sinks or sources. The solution to Eq.~\ref{master_equation} is \(\vec{p}(t) = \exp(\bigt t) \vec{p}(0)\), where \(\bigt\) is matrix notation for \(T_{ij}\),  \(T_{ii} = - \sum_{j\neq i} T_{ji}\) and  \(\vec p (t)\) is vector notation for \(p_j(t)\)~\cite{strang1988linear}. Systems represented by the master equation are completely described by the transition rate matrix, \(\bigt\), and the initial conditions \( \vec{p}(0)\). The complete solution is 
\begin{equation}\label{time_dependence}
\vec p(t) = \sum_j \vec v_j e^{\lambda_j t} \left(\mathbf{V}^{-1}_j\cdot\vec{p}(0) + a_j(t)\right),
\end{equation} 
where \( \vec v_j \) is the \(j^{th}\) eigenvector, \(\mathbf{V}^{-1}\) is the inverse of the matrix of eigenvectors, and \(a_j(t)\) is a polynomial due any eigenvalue degeneracy~\cite{dennery1967mathematics}.
Hence, characterizing the properties of \(\bigt\) also characterizes the dynamics of the system~\cite{SCHNAKENBERG:1976p1215,Hearon:1953,van2007stochastic}.  Because the time dynamics of individual modes are ultimately determined by the eigenvalues of \(\bigt\) (see Eq.~\ref{time_dependence}), we will be concerned with these eigenvalues and how they relate to oscillations.  We first explore these eigenvalues and prove how they constrain the possible oscillations to be fewer than the number of states in the system.  Second, the present work provides examples of these limits by exploring the quality of oscillations in a hypothetical stochastic clock, showing how both microscopic oscillations and macrostates are constrained by the number of states in the network. The paper concludes by proposing some experiments which may cast direct light onto the physical realization of these bounds on oscillations.

\section{The Limits on Oscillations}

To understand the oscillations in the system represented by \(\bigt\), we consider the relative contributions of different eigenmodes. From the Perron-Frobenius theorem, all eigenvalues of \(\bigt\) have nonpositive real parts, so all but the  \(\lambda=0\) equilibrium mode decay away to zero.   Eigenvectors with nonzero imaginary eigenvalues oscillate in magnitude as they decay.  As we see in Eq.~\ref{time_dependence}, after a time \((\re \lambda_i)^{-1}\), mode \(i\)'s contribution to \(\vec{p}(t)\) will have substantially diminished.  If there is an imaginary part to \(\lambda_i\),  mode \(i\) will oscillate \(|\im \lambda_i / \re \lambda_i|\) times before decaying.  Because each oscillatory mode will decay independent of the other modes, the overall quality of oscillations is given by the most oscillatory mode: 
\begin{equation}\label{oscillation_definition}
\osc = \frac12 \max_i \left | \im \lambda_i / \re \lambda_i \right |.
\end{equation} 
In a closed system, \(\osc\) is the upper bound on the total number of oscillations arising from an optimal initial state of the system.  On the other hand, in an open and homogeneously driven system, \(\osc\) describes the coherence of those oscillations, in analogy to the quality-factor of harmonic oscillators, because external driving may cause the system to oscillate indefinitely. This work establishes upper-bounds on \(\osc\) by showing the eigenvalues of \(\bigt\) only exist in specific regions of the complex plain.  

\begin{figure}[tc]
   \scalebox{0.8}{\includegraphics{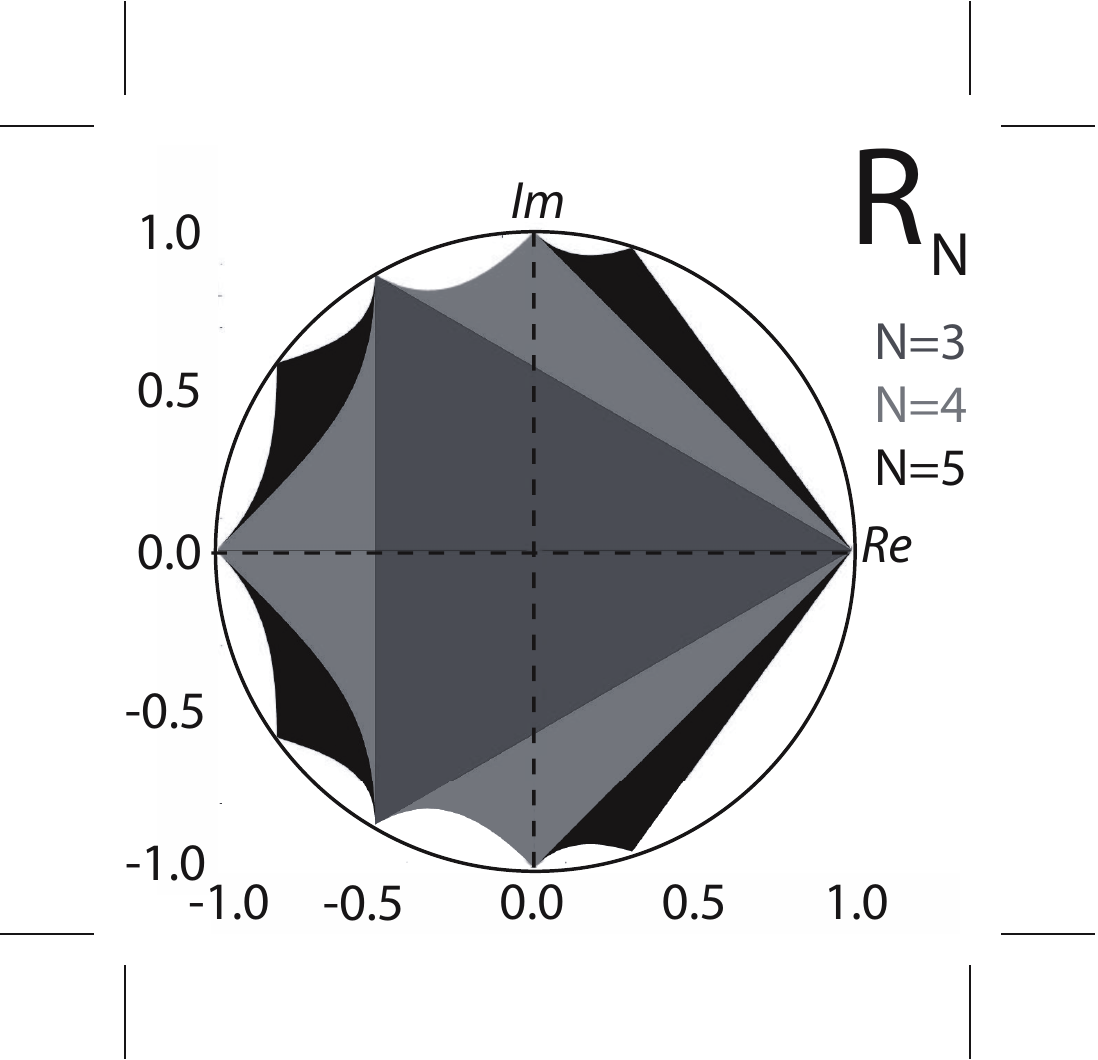} }
   \caption{The region $R_N$ contains all possible eigenvalues of $N$-dimensional stochastic matrices.  Region $R_{N+1}$ contains $R_N$.   This region is symmetric to the real axis and circumscribed by the unit circle. The curves defining each region are given by Eq.~\ref{regioncurves}, due to Karpelevich's Theorem.\label{allowedregion}}  
\end{figure}

Karpelevich's Theorem, as clarified by Ito~\cite{karpelevich1989characteristic,Ito:1997p359}, states that all possible eigenvalues of an $N$-dimensional stochastic matrix with unit spectral radius ($|\lambda_{\max}| = 1$) are contained in a bounded region,  $R_N$, on the complex plane, shown in Fig.~\ref{allowedregion}. $R_N$ intersects the unit circle at points \(\exp(2 \pi i a/b)\), where $a$ and $b$ are relatively prime and \(0 \leq a < b \leq N\).   The curve connecting points \(z=e^{2 \pi i a_1/b_1}\) and \(z=e^{2 \pi i a_2/b_2}\) is described by the parametric equation
\begin{equation}\label{regioncurves}
 z^{b_2} (z^{b1}-s)^{\lfloor N/b_1\rfloor} = z^{b_1 \lfloor N/b_1\rfloor} (1-s)^{\lfloor N /b_1 \rfloor},
 \end{equation}
where \(s\) runs over the interval \([0,1]\) and \( \lfloor x/y \rfloor \) is the integer floor of \(x/y\). For example, the curve that connects \(z=1\), corresponding to \((a_1=0,\ b_1=N)\),  with \(z=e^{2 \pi i/N}\) \((a_2=1,b_2=N)\) is \(z(s) = (e^{2 \pi i/N}-1) s +1\).

The rate matrix from Eq.~\ref{master_equation}, \(\bigt\), is not a stochastic matrix.  To preserve probability, the sum of each columns of \(\bigt\) is zero, and the diagonal elements are \(\leq 0\)\footnote{This may be obtained by letting \(p_i = (1,0,0,\dots)\), substituting \(p_i\) into Eq.~\ref{master_equation}, and solving for the condition \(\sum_i p_i = 0\).}. To transform \(\bigt\) into a stochastic matrix, denoted \(\bigt^\prime\), divide \(\bigt\) by  the sum of its largest diagonal element and largest eigenvalue then add the identity matrix.  This transformation allows us to write the eigenvalues of \(\bigt^\prime\) in terms of the eigenvalues of \(\bigt\):
\begin{equation}\label{eigenvalue_map}
\lambda^\prime_i = \frac{\lambda_i}{\max_i |\bigt_{ii}|+ \max_i |\lambda_i|} + 1.
\end{equation}
Because $\max_i \lambda_i = 0$, and all other eigenvalues of \(\bigt\) have negative real parts, the most positive eigenvalue of $\bigt^\prime$ is 1. Our normalization procedure ensures all other eigenvalues of \(\bigt^\prime\) are less than 1 and fit within the region $R_N$ on the complex plane.  Therefore,  all of the eigenvalues of $\bigt$  fit within the region \( (\max_i |\bigt_{ii}| + \max_i |\lambda_i|) \cdot (R_N -1)\).  Within this region, the maximum number of oscillations corresponds to eigenvalues along the line  $\lambda  \propto (e^{\pm 2\pi i/N} -1)$, giving 
\begin{equation}\label{central_result}
\osc_{\max} = \frac12 \left | \frac {\sin (2 \pi /N)}{\cos(2\pi/N)-1}\right| = \frac12 \cot (\pi/N)< \frac{N}{2\pi}.
\end{equation}
We can further refine the limit in Eq.~\ref{central_result} using a result from Kellogg and Stephens~\cite{KELLOGG:1978p1188}, giving
\begin{equation}\label{central_result_2}
\osc_{\max} = \frac12 \cot \frac{\pi}{\lencycle} < \frac{\lencycle}{2\pi},
\end{equation}
where \(\lencycle\) is the longest cycle in the system.  

Up to this point in our proof, we have restricted ourselves to systems without any degeneracy in the eigenvalues of \(\bigt\). With degeneracy, as shown in Eq.~\ref{time_dependence} , the time dependence of eigenvector \(j\) may pick up an extra polynomial factor, \(a_j(t)\), with degree less than the degeneracy of \(\lambda_j\), which is always less than \(N-1\). Fortuitous balancing of coefficients  could allow a \(p^{\text{th}}\)-order polynomial to add an additional \(p/2\) oscillations. Examining Eq.~\ref{time_dependence}, we see that the total maximum oscillation quality can be
\begin{equation}\label{final_limit}
\osc_{\max}  < \frac{\lencycle}{\pi} + \frac{N-1}{2} < N,
\end{equation}
where the second term is strictly due to degeneracy~\footnote{Such degeneracy will usually emerge only in hypothetical systems where rate balance perfectly.  This condition is unlikely in physical systems, but it may be relevant in abstract networks or toy systems.}


\begin{figure}[tc]	
   \scalebox{0.5}{\includegraphics{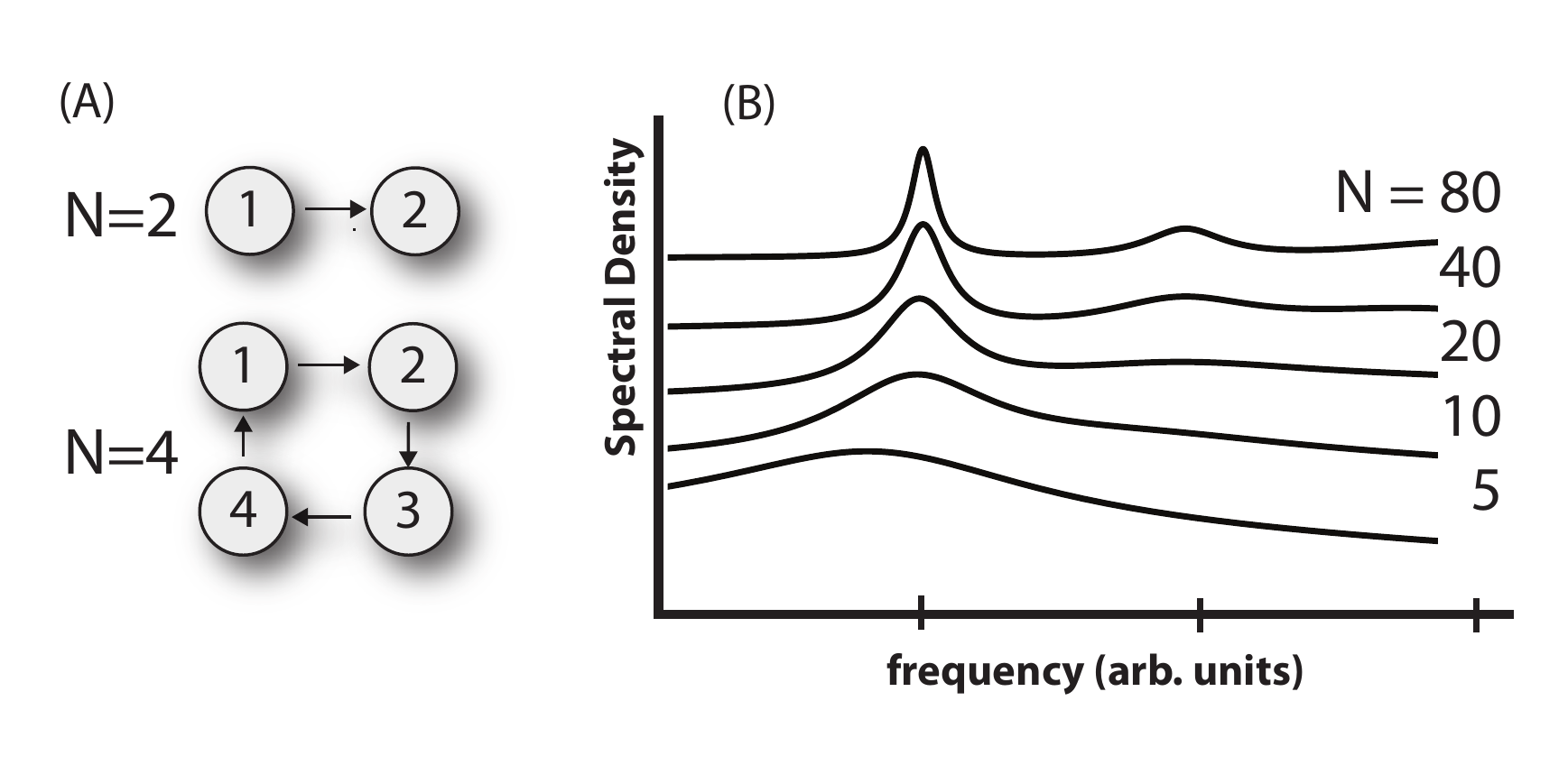}} 
   \caption{(a) When the energy landscape has barriers much larger than $k_BT$, the system will spend most of its time in the minima of the environment.  Approximating the continuous landscape by discrete states gives the familiar master equation kinetics.  Here we document two examples of systems with unidirectional transition rates.  This cyclical system  produces the maximum \(\osc\) for any given \(N\). (b)   As shown in Eq.~\ref{central_result}, a system with only two states can not coherently oscillate.  It produces only random jumps.  As the number of states in the unidirectional cycle increases (in the same family as shown in (a)), oscillations become more coherent and more persistent.  The spectral density of the unidirectional cycle shows a distinct peak which becomes sharper as \(N\) increases. The transition rates have been normalized by the number of states. The \(\osc\) of the systems are, from bottom to top: 0.75, 1.46,3.09,6.40,12.7, obtained by fitting Lorentzian functions to the peaks.\label{statespace}} 
\end{figure}

\begin{figure*}[tc]
\scalebox{1.2}{ \includegraphics{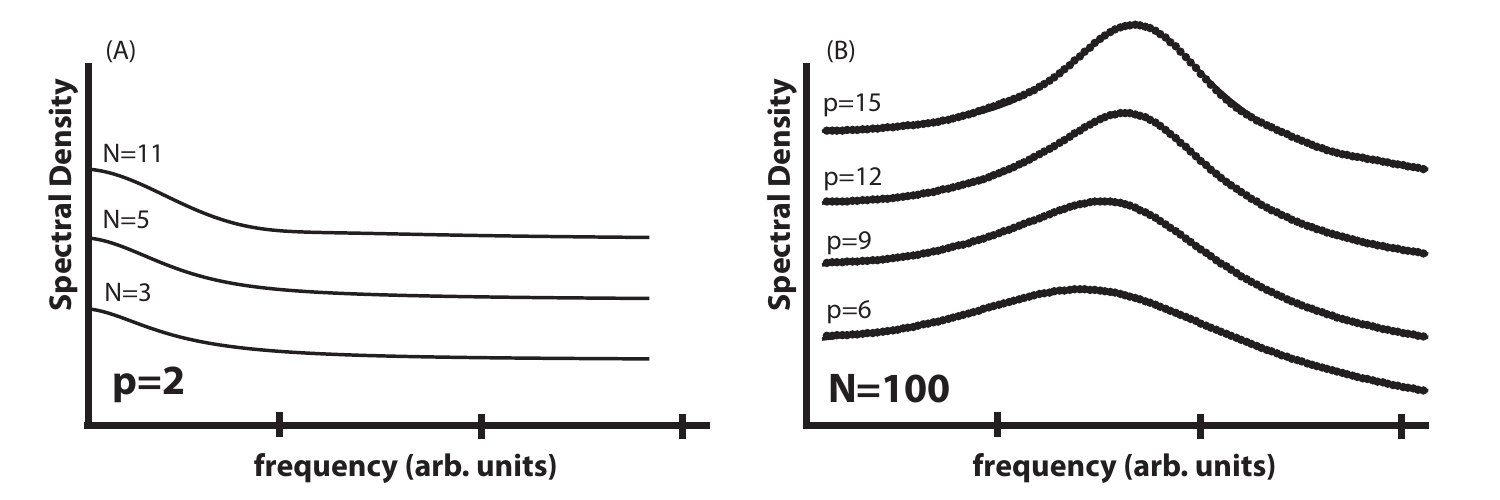}}
\caption{Although the linear set of states given by Eq.~\ref{linear_system} does not have any imaginary eigenvalues, macrostates can oscillate.  Macrostates are defined as \(\langle A \rangle  = \sum_i^N A^p_i p_i(t)\).  In this case,  \(A^p_i = \{1 \text{ if } \text{mod}_p i = 0; 0 \text{ otherwise}\}\). (a) Dynamics for different values of \(N\) for fixed \(p=2\) (inset) Identical dynamics, but with scaled rates so expected traversal times are the same. (b) Increasing \(p\) increases the number and coherence of oscillations given fixed \(N=100\), demonstrating the limit in Eq.~\ref{central_result_2}. \label{macrostates}}
\end{figure*}

\section{Oscillations in Macrostates: Chemical Clocks}
Oscillations on the discrete state-space  are only possible when the system in question violates detailed balance, which would be the case in a system which is driven.  Under detailed balance, all eigenvalues of \(\bigt\) are real~\cite{Hearon:1953}.  The microstates of this system, which is the instantaneous realization of \(\vec p(t)\), would not show any oscillatory dynamics or peaks in the spectral density, shown in Fig.~\ref{statespace}(a).  On the other hand, oscillations in macrostates, which are defined as the linear superpositions \(\langle A(t) \rangle = \sum_i A_i p_i(t)\)~\cite{Castellano:2009ce} (where \(A_i\) are some experimental observables corresponding to state $i$), do not require the underlying microstates to oscillate. The microstate probabilities only need to evolve such that  \(\langle A(t) \rangle\) oscillates.  For example, consider a hypothetical chemical clock~\footnote{The oscillating chemical clock is distinct from the traditional ``clock reaction,'' where an autocatalytic reaction causes a sudden one-time change in state on a distinct time-scale, such as the iodate-bisulphite system~\cite{scott1994Oscillations}.} described by a cycle of states~\cite{Morelli:2007p1236,Svoboda:1994p1220} 
 \begin{equation} \label{linear_system}
 s_1\rightarrow \cdots \rightarrow s_\ell \rightarrow s_1.
 \end{equation}   
   If a macroscopic clock advances one tick each time the cycle is traversed, adding states to the cycle improves the quality of the clock, as shown in Fig~\ref{statespace}(b), consistent with the limit in Eq.~\ref{final_limit}. 
   
 Now consider a more abstract implementation of a  clock cycle coupled to a fuel reservoir.  We define that the clock consumes 1 unit of fuel during the transition from $s_\ell$ back to $s_1$,  so the cycle is an open system driven by the fuel reservoir. This accounting method, in effect, unrolls the circular cycle in Eq.~\ref{linear_system} into a linear chain of microstates enumerated by the dyad \(\{ f, s_i \}\), where \(f\) is the amount of fuel remaining and \(s_i\) is the state of the clock.    The macrostate of the clock is 
\(
\langle A(t) \rangle = \sum_{i,f} A_i p(f,s_i,t) 
				= \sum_i A_i \langle p(s_i,t) \rangle_f.
\) When \(f\) is effectively infinite, the dynamics of the  closed cycle and the linear chain are equivalent.  As the system moves from one state to another, we count time by keeping track of the evolution of the macrostate  \(A(t)\).
 Therefore, the quality of oscillations in the infinite linear chain described by the dyad  \(\{ f, s_i \}\) is bounded by \(\osc_s\leq \ell\), regardless of the precise amount of fuel, demonstrating that the oscillatory limit of the macrostate is  also constrained by the size of the network representing the chemical clock in Eq.~\ref{linear_system}. 
 For example, take \(A_i = \{1 \text{ if } \text{mod}_2 i = 0; 0 \text{ otherwise}\}\), meaning $\ell = 2$.  The evolution of $\langle A(t) \rangle$ is shown in Fig.~\ref{macrostates}(a).  The total number of states, \(N\), does not effect the quality of the clock. However, if we change \(A_i\) to \[A^p_i = \{1 \text{ if } \text{mod}_p i = 0;\, 0 \text{ otherwise}\},\] Fig.~\ref{macrostates}(b) shows increasing  oscillation in \(\langle A(t) \rangle\) with increasing \(p\). That is, adding more states to the clock circuit directly increases its accuracy.

Texts exploring chemical oscillations state that nonlinearity is a requirement for oscillations.  In fact, nonlinearity is a shorthand for describing extremely large systems~\cite{van2007stochastic}.  Under conditions of detailed balance, systems must consume some sort of fuel to sustain oscillations.   If we consider the fuel-free states as being an abstract engine with \(N\) states, the combined engine-fuel system is describe by being in one of \(N\) different states and having \(f\) units of fuel remaining.  Therefore, a fuel reservoir can allow a total number of oscillations $ \approx fN/2\pi$.  Because we can approximate an extended fuel-engine network as the engine network alone combined with rates that account for the chemical potential of the fuel,  Eq.~\ref{final_limit} implies that the number of inherently unique states of the engine, absent fuel consumption, will constrain the possible regularity of reciprocal motion~\cite{Svoboda:1994p1220}.

Similarly, if the system is driven by oscillations in multiple parameters or species, we can again parameterize the state of the system based on the population of each component.  However, most chemical dynamics are modeled using continuous variables, not discrete numbers of states.    The microscopic description of the system, comprised of a discrete number of states, is connected to the  continuous mass-action approximation of chemical dynamics by a system size expansion described by Van Kampen~\cite{van2007stochastic}.  Take, as an example, the multidimensional oscillating chemical reaction called the Brusselator.  By expanding the mass-action Brusselator into a discrete state-space, the system size expansion parameter determines the length of the largest cycle~\cite{Qian:2002p1186,Gaspard:2002p1217}. 
Indeed, multiple authors have observed that the  quality  of the Brusselator limit cycle scales with system size, consistent with this work~\cite{Qian:2002p1186,Morelli:2007p1236,Xiao:2008p1203,Andrieux:2008p1214,Gonze:2002p1237,Gaspard:2002p1217}. For example, Gaspard et al. derived that the diffusion in phase-space of the Brusselator, which corresponds to amount of damping in the macroscopic observables, is inversely proportional to the size of the system~\cite{Gonze:2002p1237,Gaspard:2002p1217}. This observation is not merely coincidence, but a fundamental efficiency limit of the master equation.

\section{Conclusion}

The bounds on oscillations can play a key role in interpreting experimental observations by determining a minimum number of underlying states.  For example, the oscillation of fluorescence wavelength in fluorescent protein GFPmut2 remains unexplained~\cite{Baldini:2005p1096,Baldini:2006p524,Cannone:2007p373}.  After application of a denaturant, the ionic state of the fluorophore can switch up to \(\osc \sim 50\) times with high regularity, observed as oscillations in the emission wavelength~\cite{Baldini:2005p1096}.  Because Eq.~\ref{central_result_2} bounds the number of states involved in the oscillation  to be at least $2\pi$ times larger than \(\osc\), this predicts that the oscillations are driven by large-scale rearrangement of the numerous hydrogen bonds in the \(\beta\)-barrel, not merely exchange between the few amino acids directly connected to the fluorophore.  If the protein were to be mutated to alter the number of bonds in the \(\beta\)-barrel, we predict that we should see a corresponding alteration in the number and quality of observed oscillations.  

The quality bounds proved hear are universal.  Because the master equation is used in nearly every branch of science, the dynamics being modeled need not be physical~\cite{Castellano:2009ce}.  For example, it could be money held by a bank~\cite{Helbing:2004p530}, packets of data on the internet~\cite{Bridgewater:2005p1227}, agents traversing a network~\cite{Shreim:2007p683}, or the populations in an ecosystem~\cite{Mckane:2005p1239}. The oscillation limit could also be probed experimentally with sculpted landscapes using optical tweezers~\cite{Wu:2009p1244}.  As a probe bead jumps from trap to trap, the energy landscape in unoccupied traps is sculpted to simulate an arbitrarily large designer network of discrete states. Ultimately, the current results are a fresh approach to analyzing the dynamics of discrete systems, and it serves as a new design principle for those seeking to engineer oscillations. 

 This efficiency limit of oscillations has obvious implications on how well a high-dimensional system can be numerically approximated by a smaller system.  The approximation will only be successful if the relevant  eigenvalues of the larger system lie within the allowed region of the smaller system. However, the inverse stochastic eigenvalue problem has not yet been solved, so we can not know \textit{a priori} if a stochastic matrix exists for a given set of eigenvalues, even if they all reside within the allowed region~\cite{Egleston:2004p1196}. This fact prevents us from constructing the opposite bounds, the conditions for a \textit{minimum} number of oscillations.  Hopefully, future results will further constrain the present bounds, and we may gain deeper insight into the necessary conditions for creating oscillations.

\begin{acknowledgments}
I would like to thank Scott Fraser, and Rudy Marcus for insightful discussions and support. Many thanks to Nima Ghaderi, Hideo Mabuchi and David Politzer.  Special note to Benjamin Rahn  for inspiration.  This was supported by the Beckman Institute (Biological Imaging Center), the Office of Naval Research (N00014-07-1-0442), and the
National Science Foundation (CHE-0848178).
\end{acknowledgments}

\bibliography{oscillationsfull}

\end{document}